\documentclass[conference, final]{IEEEtran}

\IEEEoverridecommandlockouts

\usepackage{cite}
\usepackage{amsmath,amssymb,amsfonts}
\usepackage{algorithm}
\usepackage{algorithmicx}
\usepackage{graphicx}
\usepackage{textcomp}
\usepackage{xcolor}
\usepackage{bm}
\usepackage{multirow}
\usepackage{array}
\usepackage[noend]{algpseudocode}
\usepackage{subcaption}
\usepackage{url}
\usepackage{color}

\makeatletter
\newcommand\fs@spaceruled{\def\@fs@cfont{\bfseries}\let\@fs@capt\floatc@ruled
  \def\@fs@pre{\vspace{.65\baselineskip}\hrule height.8pt depth0pt \kern2pt}%
  \def\@fs@post{\kern2pt\hrule\relax}%
  \def\@fs@mid{\kern2pt\hrule\kern2pt}%
  \let\@fs@iftopcapt\iftrue}
\makeatother

\def\BibTeX{{\rm B\kern-.05em{\sc i\kern-.025em b}\kern-.08em
    T\kern-.1667em\lower.7ex\hbox{E}\kern-.125emX}}

\newcolumntype{A}{>{\centering\arraybackslash}m{0.08\textwidth}}
\newcolumntype{D}{>{\centering\arraybackslash}m{0.09\textwidth}}
\newcolumntype{B}{>{\centering\arraybackslash}m{0.15\textwidth}}
\newcolumntype{C}{>{\centering\arraybackslash}m{0.25\textwidth}}
\newcolumntype{E}{>{\centering\arraybackslash}m{0.25\textwidth}}

\newcommand{\algparbox}[2]{\parbox[t]{\dimexpr\linewidth-#1}{#2\strut}}
\algrenewcommand\algorithmicindent{1em}%

\begin{document}

\title{
R-Learning-Based Admission Control for Service Federation in Multi-domain 5G Networks

\thanks{{This  work  has  been  partially  funded  by  the  MINECO  grant TEC2017-88373-R (5G-REFINE), the EC  H2020  5Growth  Project  (grant  no.  856709),  and Generalitat de Catalunya grant 2017 SGR 1195.}}

}

\author{
\IEEEauthorblockN{
	Bahador Bakhshi,
	Josep Mangues-Bafalluy, 
	Jorge Baranda
}
\IEEEauthorblockA{Centre Tecnologic de Telecomunicacions de Catalunya (CTTC), Spain}
}

\maketitle

\begin{abstract}
Network service federation in 5G/B5G networks enables service providers to extend 
service offering by collaborating with peering providers. Realizing this vision requires interoperability 
among providers towards end-to-end service orchestration across multiple administrative domains. Smart admission control is fundamental to make such extended offering
profitable.
{Without prior knowledge of service requests,}
the admission controller (AC) either determines the domain to deploy the demand or rejects it
to maximize the long-term average profit. In this paper, we first obtain the optimal AC policy by formulating the problem as a Markov decision process, which is solved through the policy iteration method. {This provides the  \textit{theoretical} performance bound} under the assumption of known arrival and departure rates of demands. 
{Then, for practical solutions to be }deployed in real systems, 
where the rates are not known, we apply the Q-Learning and R-Learning algorithms to approximate the optimal policy. The extensive simulation results show that learning approaches outperform the greedy policy and are capable of getting close to optimal performance.
More specifically, R-learning always outperformed the rest of practical solutions and achieved an optimality gap of 3-5\% independent of the system configuration, while Q-Learning showed lower performance and depended on discount factor tuning.

\end{abstract}


\section{Introduction}
5G and beyond networks are expected to cost-effectively provide services with diverse quality requirements. To this end, enabler technologies including Network Function Virtualization (NFV) and Software Defined Networking (SDN) 
as well as  architectural principles like network slicing, and multi-domain orchestration are incorporated in the architecture of the networks  \cite{alliance20175g, virtualisation2018release}.
The variety of services and stakeholders needs the definition of management and orchestration architectures
for the cooperation of stakeholders
that is required towards end-to-end multi-domain orchestration of such services. One clear example is service federation \cite{valcarenghi2018framework, hortiguela2020realizing}.

Multi-domain orchestration enables the service provider, as an administrative domain, to collaborate with other domains for service provisioning in a federated environment \cite{hortiguela2020realizing}. The federation contract between the domains
provides extra resources for the consumer domain at the cost of the federation. These resources are used to deploy (segments of) the requested network services, i.e., a network slice composed of a set of Virtual Network Functions (VNFs) and virtual interconnecting links, in other domains
for different reasons such as the lack of sufficient local resources, load-balancing, and cost-efficiency. 

The Admission Controller (AC) 
is the highest-level resource manager. 
For a given service demand, it either determines the domain to deploy the demand or rejects it. In this paper, we consider business profit as the objective of the admission control, and assume that there is only one provider domain with a limited reserved quota for service federation which is agreed in the federation contract. 
Under these assumptions, the AC should decide where to deploy the requested services
to maximize the profit
without knowing the future demands. This problem is referred to as the Admission Control for Service Federation (ACSF). 

This problem is challenging, and trivial approaches like greedily accepting every demand do not provide the optimal solution, as shown in the simulation results.
The AC should manage the local resources of the consumer domain and the federation quota of the provider domain to maximize the profit. While several AC algorithms in 5G networks have already been proposed \cite{ojijo2020survey}, they consider single domain networks; and consequently are not directly applicable to the ACSF problem. This is the research gap that we aim to address.

Recently, AI/ML approaches have been extensively used 
in communication networks \cite{morocho2019machine}.
{ACFS is an instance of the problem of sequential decision making under uncertainty, which can be efficiently approached by Reinforcement Learning (RL) solutions where an agent learns the policy via interaction with the environment \cite{sutton2018reinforcement}}.
The RL based admission control solutions have already been developed in other contexts rather than the multi-domain service federation problem \cite{tong2000adaptive,liu2005self, wu2001admission, caballero2018network}. In this paper, we utilize a special category of reinforcement learning, named \textit{average reward} learning, to approximate the optimal policy that maximizes the average profit of the consumer domain. More specifically, we make the following contributions to the ACSF problem:
$i$) The ACSF problem is formulated as a Markov Decision Process (MDP), which is solved by the Policy Iteration Dynamic Programming (DP) method to obtain the theoretical performance bound.
$ii$) An average reward based learning algorithm is developed to approximate the optimal policy.
$iii$) We show that the commonly used Q-Learning algorithm does not perform well in ACSF due to dependency on the discount factor.

The remainder of this paper is
as follows. In Section \ref{sec_related}, we review the related works. 
The system models and the MDP formulation 
are presented in Section \ref{sec_model}. The model is solved by the Policy Iteration (PI) algorithm in Section \ref{sec_PI}. We present the learning approaches in Section \ref{sec_QLRL} and evaluate them 
in Section \ref{sec_sim}. Section \ref{sec_summary} concludes this paper. 

\section{Related Work}
\label{sec_related}

Given the wide variety of stakeholders of future networks, service federation is expected to become a relevant component of 5G and beyond network architecture \cite{ alliance20175g}, 
since it allows deploying complex services in multiple administrative domains, whilst enabling end-to-end orchestration. At a theoretical level, the problem of service federation is formulated as an Integer Linear Programming model in \cite{sun2018service}, which is extended 
to consider energy efficiency \cite{sun2019energy}. 
From an architectural point of view, preliminary ideas of service federation were presented in \cite{valcarenghi2018framework} and \cite{li2018service}. The full-fledged architecture enabling service federation was 
developed
in the 5G-Transformer \cite{5GTrans} platform, which is capable of deploying composite NFV network services spanning across multiple domains
and realizes the federation vision by designing the high-level concepts presented in ETSI specifications \cite{hortiguela2020realizing}. However, neither the theoretical nor the architectural research works address the admission control problem in service federation; i.e., they assume that the service has already been accepted and attempt to efficiently deploy it.

Using reinforcement learning for admission control has been the topic of several previous works. Adaptive call admission control (CAC) in multimedia networks using RL  was studied in \cite{tong2000adaptive}.  In the case of links with variable capacity, the CAC problem using RL was investigated in \cite{pietrabissa2011reinforcement}.
In \cite{liu2005self}, the authors utilized  RL for CAC in CDMA networks. In \cite{wu2001admission}, admission control in cellular networks, and in \cite{han2018markov} network slice admission control were formulated as MDP. The network slice admission control in single domain networks was studied in \cite{raza2018slice, caballero2018network}.
While these works approach the AC problem using RL, they cannot  be applied directly to the ACSF problem since they consider single domain networks. 

In \cite{antevski2020q}, an initial attempt to apply Q-Learning to the service federation problem was presented. In this paper, we go beyond
by formulating the problem as an MDP to obtain the optimal solution and we propose a new average-reward learning algorithm that outperforms Q-Learning under all evaluated scenarios and whose performance is not as sensitive to parameter tuning (e.g., discount factor) as Q-learning.

\section{System Model and Problem Formulation}
\label{sec_model}

\subsection{Assumptions and System Model}
In this paper, service federation between a consumer domain and a provider domain is considered, where as mentioned, each requested service is a network slice composed of a set of VNFs and virtual interconnecting links. It is assumed that in the consumer domain, only one type of the resources (e.g., CPU) has a limited capacity $LC$. The domain offers a set of services, denoted by $\mathcal{I}$. Each service type $i \in \mathcal{I}$ is specified by $(w_{i},r_{i})$ in the service catalog, where $w_{i}$ is the total amount of required resources (e.g., the total number of CPU cores) by all the VNFs of the service type, and $r_{i}$ is the revenue that the provider gains if it accepts a demand of this type.

For service federation, the provider domain reserved a resource quota with capacity $PC$ units, which is agreed in the federation contract. 
The contract also specifies the federation cost $\phi_{i}$, which is the cost of deploying an instance of service type $i$ in the provider domain that is paid by the consumer domain. Demands beyond the capacity $PC$ will be rejected.

Over time, customers request instances of the services. The load of service type $i$ corresponds to a traffic class $(\lambda_{i}, \mu_{i}, w_{i}, r_{i}, \phi_{i})$ where $\lambda_{i}$ and $\mu_{i}$ are, respectively, the arrival and departure rates of the demands of the type, which are assumed to be a Poisson process. 
Accordingly, each demand $\delta_{i}$ of class $i$ is determined by $(\tau^{s}_{\delta}, \tau^{e}_{\delta}, w_{i}, r_{i}, \phi_{i})$ where the $\tau^{s}_{\delta}$ and $\tau^{e}_{\delta}$ are the start and end time of the demand, which are respectively specified by $\lambda_{i}$ and $\mu_{i}$.

Upon arrival of $\delta_{i}$, the AC should determine in which domain to deploy the demand or to reject it. To accept the demand in the consumer domain, $w_{i}$ units of the \textit{local} resources are allocated to the demand and $r_i$ units of money is earned. In the case of the federation, the demand is deployed in the provider domain; so, no resource is consumed in the local domain and the profit is $r_{i}-\phi_{i}$. If the demand is rejected, no resource is used and no revenue is earned.

In ACSF, we assume there is a set $\mathcal{D}$ of demands, which arrive one-by-one. 
At the arrival time of a demand, AC is not aware of the future demands and also does not know the expected life-time of the given demand. The objective is to find an AC policy that maximizes the average profit, which is 
\begin{equation}
\label{eq_profit}
\frac{1}{|\mathcal{D}|}\sum_{i \in \mathcal{I}}\Big(\sum_{\delta_{i} \in \mathcal{L}}  r_{i} + \sum_{\delta_{i} \in \mathcal{F}} (r_{i} - \phi_{i})\Big),
\end{equation}   
where $\mathcal{L}$ and $\mathcal{F}$ are, respectively, the set of the demands that are deployed in the local consumer domain and in the federated provider domain. 


\subsection{The MDP Model of ACSF}

\begin{table*} 
\begin{center}
\centering
\small\addtolength{\tabcolsep}{0pt}
\caption{The Transition Probabilities and Rewards of the MDP}
\label{table_mdp}
\scalebox{0.8}{%
\begin{tabular}{|D|A|C|B|E|A|}
\hline
Current state $s$ & Action in state $s$ & Next event after this action	& Next state $s'$ & Transition probability $P_{a}(s,s')$	& Reward $R_{a}(s,s')$ \\
\hline
\hline
$(\bm{l}, \bm{f}, \bm{e}_{i})$ & \textsf{reject} & Arrival of a demand of class $j$ & $(\bm{l}, \bm{f}, \bm{e}_{j})$ & $\frac{\lambda_{j}}{\Lambda(s') + M(s')}$ & 0 \\
\hline		

$(\bm{l}, \bm{f}, \bm{e}_{i})$ & \textsf{reject} & Departure of a demand of class $j$ & $(\bm{l}, \bm{f}, -\bm{e}_{j})$ & $\frac{(\bm{l}'[j] + \bm{f}'[j])\mu_{j}}{\Lambda(s') + M(s')}$ & 0 \\
\hline		

$(\bm{l}, \bm{f}, \bm{e}_{i})$ & \textsf{accept} & Arrival of a demand of class $j$ & $(\bm{l}+\bm{e}_i, \bm{f}, \bm{e}_{j})$ & $\frac{\lambda_j}{\Lambda(s') + M(s')}$ & $r_{i}$ \\
\hline		

$(\bm{l}, \bm{f}, \bm{e}_{i})$ & \textsf{accept} & Departure of a demand of class $j$ & $(\bm{l}+\bm{e}_{i}, \bm{f}, -\bm{e}_{j})$ & $\frac{(\bm{l}'[j] + \bm{f}'[j])\mu_{j}}{\Lambda(s') + M(s')}$ & $r_{i}$ \\
\hline		

$(\bm{l}, \bm{f}, \bm{e}_{i})$ & \textsf{federate} & Arrival of a demand of class $j$ & $(\bm{l}, \bm{f}+\bm{e}_i, \bm{e}_{j})$ & $\frac{\lambda_j}{\Lambda(s') + M(s')}$ & $r_{i} - \phi_{i}$ \\
\hline		

$(\bm{l}, \bm{f}, \bm{e}_{i})$ & \textsf{federate} & Departure of a demand of class $j$ & $(\bm{l}, \bm{f}+\bm{e}_{i}, -\bm{e}_{j})$ & $\frac{(\bm{l}'[j] + \bm{f}'[j])\mu_{j}}{\Lambda(s') + M(s')}$ & $r_{i} - \phi_{i}$ \\
\hline		

$(\bm{l}, \bm{f}, -\bm{e}_{i})$ & \textsf{no\_action} & Arrival of a demand of class $j$ & $(\bm{l}-\bm{e}_{i}, \bm{f}, \bm{e}_{j})$ & $\frac{\bm{l}[i]}{\bm{l}[i] + \bm{f}[i]} \times  \frac{\lambda_{j}}{\Lambda(s') + M(s')}$ & 0 \\
\hline 

$(\bm{l}, \bm{f}, -\bm{e}_{i})$ & \textsf{no\_action} & Arrival of a demand of class $j$ & $(\bm{l}, \bm{f}-\bm{e}_{i}, \bm{e}_{j})$ & $\frac{\bm{f}[i]}{\bm{l}[i] + \bm{f}[i]} \times  \frac{\lambda_{j}}{\Lambda(s') + M(s')}$ & 0 \\
\hline 

$(\bm{l}, \bm{f}, -\bm{e}_{i})$ & \textsf{no\_action} & Departure of a demand of class $j$ & $(\bm{l}-\bm{e}_{i}, \bm{f}, -\bm{e}_{j})$ & $\frac{\bm{l}[i]}{\bm{l}[i] + \bm{f}[i]} \times  \frac{(\bm{l}'[j] + \bm{f}'[j])\mu_{j}}{\Lambda(s') + M(s')}$ & 0 \\
\hline		

$(\bm{l}, \bm{f}, -\bm{e}_{i})$ & \textsf{no\_action} & Departure of a demand of class $j$ & $(\bm{l}, \bm{f}-\bm{e}_{i}, -\bm{e}_{j})$ & $\frac{\bm{f}[i]}{\bm{l}[i] + \bm{f}[i]} \times  \frac{(\bm{l}'[j] + \bm{f}'[j])\mu_{j}}{\Lambda(s') + M(s')}$ & 0 \\
\hline 

\end{tabular}	
}		
\end{center}
\end{table*}

In this section, under the assumption of knowing  $\lambda_{i}$ and $\mu_{i}$ 
$\forall i \in \mathcal{I}$, the ACSF problem is formulated as a Markov decision process.
An MDP is a 4-tuple $\big(\mathcal{S}$, $\mathcal{A}$, $P_{a} (s,s')$, $R_{a}(s,s')\big)$ where $\mathcal{S}$ is the set of the states, $\mathcal{A}$ is the set of actions, $P_{a}(s,s')$ is the transition probability from state $s$ to state $s'$ if the agent takes action $a$ in the state $s$, and $R_{a}(s,s')$ is the reward of the action $a$ in state $s$ that leads to transition to state $s'$.

In ACSF, the state of the environment is defined as
$s=(\bm{l}, \bm{f}, \bm{d})$,
where $\bm{l}$ and $\bm{f}$ are vectors wherein the $i$-th element is the number of active (alive) demands of the class $i$ in the local consumer domain and in the provider domain, respectively. Similar to \cite{wu2001admission}, $\bm{d}$ is a vector whose $i$-th element is $+1$ if a demand of class $i$ arrives, it is $-1$ if a demand of class $i$ departs the network, and is 0 otherwise.
Let $\mathcal{S}^{+} = \{s \, |\, \exists i  \text{ s.t. }  \bm{d}[i]= +1 \}$ and similarly $\mathcal{S}^{-}= \{s \, | \, \exists i  \text{ s.t. } \bm{d}[i]= -1\}$; we have these constrains on the states to make the problem tractable:
$i$) $\mathcal{S}= \mathcal{S}^{+} \cup \mathcal{S}^{-}$  and $\mathcal{S}^{+} \cap \mathcal{S}^{-} = \{\}$. 
$ii$) State $(\bm{l},\bm{f},\bm{0})$ is an invalid state as there is not any arrival or departure events.
$iii$) State $(\bm{l},\bm{f},\bm{d})$ where $\exists i,j \text{ s.t. } \bm{d}[i] \neq 0 \text{ and } \bm{d}[j] \neq 0$ is an invalid state as two events cannot occur at the same time\footnote{Please note that the AC
only takes an action in states $s \in \mathcal{S}^{+}$; however, we consider states $s \in \mathcal{S}^{-}$ in the MDP as these states correspond to the departure of demands wherein the capacity of the resources changes. These states are used to facilitate deriving the transition probabilities.}.

$\mathcal{A}(s)$ is the set of 
\textit{valid} actions in state $s$. Define $\widetilde{LC}$ and $\widetilde{PC}$ as the current \textit{available} capacity of resources in the consumer and provider domains, respectively. For $s \in \mathcal{S}^{+}$ where $\exists i \text{ s.t. } \bm{d}[i] = +1$,
$\mathcal{A}(s)$ includes $i$) {\small \textsf{reject}}, $ii$) {\small \textsf{accept}} only if $\widetilde{LC} \geq w_{i}$, and $iii$) {\small \textsf{federate}} only if $\widetilde{PC} \geq w_{i}$. However, if $s \in \mathcal{S}^{-}$ then the only valid action is an artificial action, named {\small \textsf{no\_action}}, since in this case, the agent does not do any action and a demand departs the system.

As mentioned above, it is assumed that parameters $\lambda_{i}$ and $\mu_{i}$ are known and determine the transition \textit{rates}. In MDP, the transition \textit{probabilities} should be derived. According to the theory of competing exponentials, the probability of a transition in a state equals to the rate of the transition divided by the total transition rates in the state. The total transitions rates of state $s=(\bm{l}, \bm{f}, \bm{d})$ is $\Lambda(s)+ M(s)$ where $\Lambda(s)= \sum_{i \in \mathcal{I}} \lambda_{i}$ is the total arrival rate of demands and $M(s)=\sum_{i \in \mathcal{I}}(\bm{l}[i]+\bm{f}[i])\mu_i$ is the total departure rate of the demands in the state.

We assume the transition from $s$ to $s'$ occurs in two steps. At first, the agent makes a decision that takes place immediately in the system, i.e., $\bm{l}$ or $\bm{f}$ changes to $\bm{l}'$ or $\bm{f}'$
 \textit{before} occurring the next arrival/departure event. Then, in the second step, the environment brings up a new event, i.e., it changes $\bm{d}$ to $\bm{d}'$, and consequently the system goes from $s=(\bm{l}, \bm{f}, \bm{d})$ to $s'=(\bm{l}', \bm{f}', \bm{d}')$. 
Therefore, to compute the transition probability from $s$ to $s'$, the total transition rates of state $s'$ should be used 
as the rate of the next event depends on $\bm{l}'$ and $\bm{f}'$. 

Let $\bm{e}_{i}$ be a vector with a 1 in the $i$-th element and 0 in the others. The transition probabilities and corresponding rewards are shown in Table \ref{table_mdp}. They are computed as follows. First, we apply the action in state $s$ that determines $\bm{l}'$ and $\bm{f}'$, then we utilize the competing exponentials theorem for the given next event. For example, the first row in the table is the case that a new demand of class $i$ arrives, so the state is $s=(\bm{l}, \bm{f}, \bm{e}_{i})$. The agent decides to reject it; hence, the reward is 0, $\bm{l}' =\bm{l}$, and $\bm{f}'=\bm{f}$. The next event after the action is the arrival of a demand from class $j$, so the next state $s'$ is $(\bm{l}, \bm{f}, \bm{e}_{j})$ with probability of $\frac{\lambda_{j}}{\Lambda(s') + M(s')}$. If the next event is the departure of a demand of class $j$, the next state will be $s'=(\bm{l}, \bm{f}, -\bm{e}_{j})$; the rate of this event is $(\bm{l}'[j] + \bm{f}'[j])\mu_{j}$, which is shown in the second row of the table. 


\section{Optimal Policy by Dynamic Programming}
\label{sec_PI}
Given the MDP model, Dynamic Programming (DP) algorithms, e.g., the Policy Iteration (PI) method, can solve it and find the optimal policy \cite{sutton2018reinforcement}. The method is depicted in Algorithm \ref{alg_PI}, where the parameter $\theta$ determines the precision and $\gamma$ is the reward discount factor. In the policy iteration loop, the given policy $\pi$ is evaluated by updating the state values $V(s)$ using the transition probabilities and rewards until the desired precision is achieved; then in the policy improvement loop, for each state, the old action $\bar{a}$ is compared against the new action obtained from the updated $V(s)$; if they are not the same, then these two loops are iterated. 

To solve the ACSF problem using the PI algorithm, an important issue needs to be addressed properly. The optimal policy found by this algorithm in fact optimizes $V(s)$, which is the \textit{discounted} state values. Therefore, the policy is optimal with respect to the discount factor $\gamma$. Different values of $\gamma$ can/may lead to different policies; e.g., $\gamma = 0$ implies that the policy only aims to maximize the one-step/immediate reward $R_a(s,s')$, which is exactly the greedy policy does.

In ACSF, the objective is to maximize the average profit defined by \eqref{eq_profit}, which is, indeed,  the \textit{average reward} rather than the discounted reward optimized by the PI algorithm. While some DP methods have been proposed in the literature to find the optimal average reward policy, it is also known that by $\gamma \rightarrow 1$, maximizing the discounted reward approximates the average reward \cite{schwartz1993reinforcement}. So, in the ACSF problem, we set $\gamma \approx 1$ to approximate the optimal policy.

While the PI algorithm can find the optimal policy, in most practical circumstances, the arrival and departure rates of the demands are not known; moreover, the number of states exponentially grows in terms of $|\mathcal{I}|$, $LC$, and $PC$. Hence, the DP methods are not practical solutions. We will use them to obtain the theoretical performance bound for evaluating the practical solutions presented in the following sections. 

\floatstyle{spaceruled}
\restylefloat{algorithm}
\begin{algorithm}
\caption{PI($\mathcal{S}$, $\mathcal{A}$, $P_{a}$, $R_{a}, \theta, \gamma$)}
\label{alg_PI}
\begin{small}
\begin{algorithmic}[1]
	\State Arbitrarily initialize $V(s) \in \mathbb{R}$ and $\pi(s) \in \mathcal{A}(s)$   $\forall s \in \mathcal{S}$
	\While {$\pi$ is not stable}
		\While {$\Delta > \theta$} \Comment{The policy evaluation loop}
			\State $\Delta \gets 0$
			\For {$\forall s \in \mathcal{S}$}
				\State $v \gets V(s)$
				\State $a \gets \pi(s)$
				\State $V(s) \gets \sum_{s'} P_a(s,s')\big(R_a(s,s')+\gamma V(s')\big)$
				\State $\Delta \gets \text{max}(\Delta, |V(s)- v|)$
			\EndFor
		\EndWhile 
		
		\For {$\forall s \in \mathcal{S}$} \Comment{The policy improvement loop}
			\State $\bar{a} \gets \pi(s)$
			\State $\pi(s) \gets \text{argmax}_{a} \sum_{s'} P_a(s,s')\big(R_a(s,s')+\gamma V(s')\big) $
			\If {$\bar{a} \neq \pi(s)$}
				\State $stable \gets false$
			\EndIf
		\EndFor
	\EndWhile
	\State \Return $\pi$
\end{algorithmic}
\end{small}	
\end{algorithm}

\section{Practical Solution via Learning}
\label{sec_QLRL}
In this section, RL approaches are applied to the ACSF problem to deal with the issues of the DP solutions. These approaches instead of finding the optimal policy by exploiting the transition probabilities, \textit{learn} the policy via interaction with the environment over time; therefore, they need neither the transition probabilities nor enumerating all the possible states.

\subsection{Q-Learning}
\label{sec_QL}
Q-Learning is one of the well-known RL algorithms that use the concept of \textit{temporal difference} to iteratively solve the Bellman optimality equations. Details of the algorithm are depicted in Algorithm \ref{alg_QL} \cite{sutton2018reinforcement}. It maintains a table $Q$ of values of each action in each state, denoted by $Q[s,a]$. At state $s$, the agent selects an action which is determined by the values $Q[s,.]$ and the exploration strategy. Then, it observes the next state $s'$, gets reward $R_a(s,s')$ from the environment, and consequently  updates the $Q[s,a]$ as follows
\begin{equation*}
Q[s,a] = (1 - \alpha)Q[s,a] + \alpha \Big(R_{a}(s,s') + \gamma \max_{a'} Q[s',a']\Big), 
\end{equation*}   
where $\alpha$ is the learning rate.  These interactions take place in $n$ learning episodes with $m$ number of demands per episode.

The general Q-Learning algorithm is customized for the ACSF problem as follows. Action in each state is chosen by the $\epsilon$-greedy strategy \cite{sutton2018reinforcement}.
At the beginning of each episode, the values of parameters $\alpha$ and $\epsilon$ are decayed to rely more on the learned $Q$ values over time. In the beginning, the large value of $\alpha$ causes the agent to learn faster and the large value of $\epsilon$ allows it to explore more. But later, decreasing these parameters in subsequent episodes forces the agent to pay more attention to the $Q$ values that it has learned. 

\begin{algorithm}
	\caption{Q-Learning($n$, $m$, $\alpha$, $\gamma$, $\epsilon$)}
	\label{alg_QL}
	\begin{small}
\begin{algorithmic}[1]
	\State Arbitrarily initialize $Q[s,a] \in \mathbb{R}$  $\forall s \in \mathcal{S}$, $\forall a \in \mathcal{A}(s)$
	\For {$n$ times}
		\State $\alpha \gets 0.99 \alpha$, $\epsilon \gets 0.99 \epsilon$
		\State $s \gets $ environment state $(\bm{0}, \bm{0}, \bm{d})$
		\For {$m$ times}
			\State $a \gets $ action from $\mathcal{A}(s)$ by $\epsilon$-greedy strategy
			\State Action $a$ is performed by the environment
			\State $s', R_{a}(s,s') \gets$ next state and reward from the environment
			\State {\small \algparbox{1.5em}{$Q[s,a] \gets (1 - \alpha)Q[s,a] + \alpha \Big(R_{a}(s,s') + \gamma \max_{a'} Q[s',a']\Big)$}}
			\State $s \gets s'$
		\EndFor
	\EndFor
	\State $\pi(s) \gets \text{argmax}_{a} Q[s,a]$ $\forall s \in \mathcal{S}$
	\State \Return $\pi$
\end{algorithmic}
\end{small}	
\end{algorithm}

Unlike the parameters $\alpha$ and $\epsilon$, the proper setting of $\gamma$ is not straightforward. Similar to PI, $\gamma = 0$ turns Q-Learning to the greedy policy; however, $\gamma \approx 1$ doesn't work well because of \textit{bootstrapping} { where the value of the next state is overestimated as $\max_{a'} Q[s',a']$. While $\gamma > 0$ allows the agent to consider expected future rewards in decision making, $\gamma \rightarrow 1$ in combination with bootstrapping can cause the value of $\gamma \max_{a'} Q[s',a']$ surpasses $R_{a}(s,s')$ in the value update equation;} and consequently, the agent underestimates the importance of the immediate rewards.
In ACSF, it implies that while $R_{\textsf{accept}}(s,s') > R_{\textsf{federate}}(s,s')$, the agent may prefer {\small \textsf{federate}}
{instead of {\small \textsf{accept}}. This is problematic in the case of $LC \gg \sum_{i \in I} (\lambda_{i}/\mu_{i})w_{i}$ where
the optimal action is {\small \textsf{accept}} $\forall s \in \mathcal{S}^{+}$; but the Q-Learning policy incorrectly selects {\small \textsf{federate}} for some states that leads to a sub-optimal policy.} The effect of $\gamma$ is evaluated in Section \ref{sec_sim}.

\subsection{R-Learning}

In this section, to resolve the issue of the discount factor, which is needed in Q-Learning, we apply \textit{average reward} RL to the ACSF problem, where the agent directly maximizes the average reward instead of the discounted reward \cite{dewanto2020average}.

The R-Learning algorithm is one of the average reward RL solutions \cite{schwartz1993reinforcement}. The details of the algorithm are shown in Algorithm \ref{alg_RL}. Due to the similarities between the Q-Learning and R-Learning algorithms, we omit the explanation of the common steps and only emphasize the differences. Contrary to the Q-Learning, the state-action values, $Q[s,a]$, are not the expected discounted reward. The key idea of the R-Learning algorithm is that in the infinite horizon and ergodic MDPs, the average reward, which is denoted by $\rho$ in the algorithm, is independent of the state. Therefore, the algorithm  by $Q[s,a]$ keeps tracking the difference between $\rho$ and the expected average reward of action $a$ in state $s$. The $Q[s,a]$ value is updated according to the difference between the expected average reward $\rho$ and the immediate reward $R_{a}(s,s')$ and also the value of the next state as follows:
\begin{equation*}
Q[s,a] \gets (1 - \alpha)Q[s,a] + \alpha \Big(\big(R_{a}(s,s') - \rho\big) +  \max_{a'} Q[s',a'] \Big)
\end{equation*}

Since the average reward is not known at the beginning, the algorithm also learns it. As seen in line 12, $\rho$ is updated by the learning rate $\beta$. The conditional update is to avoid the skews made due to randomness of the exploration strategy \cite{schwartz1993reinforcement}.

\begin{algorithm}
	\caption{R-Learning($n$, $m$, $\alpha$, $\beta $, $\epsilon$)}
	\label{alg_RL}
	\begin{small}
\begin{algorithmic}[1]
	\State Arbitrarily initialize $Q[s,a] \in \mathbb{R}$  $\forall s \in \mathcal{S}$, $\forall a \in \mathcal{A}(s)$, $\rho \gets 0$
	\For {$n$ times}
		\State $\alpha \gets 0.99 \alpha$, $\epsilon \gets 0.99 \epsilon$,  $\beta \gets 0.99 \beta$
		\State $s \gets $ environment state $(\bm{0}, \bm{0}, \bm{d})$
		\For {$m$ times}
			\State $a \gets $ action from $\mathcal{A}(s)$ by $\epsilon$-greedy strategy
			\State Action $a$ is performed by the environment
			\State $s', R_{a}(s,s') \gets$ next state and reward from the environment
			\State \algparbox{5.5em}{$Q[s,a] \gets (1 - \alpha)Q[s,a] + \alpha \Big(\big(R_{a}(s,s') - \rho\big) +  \max_{a'} Q[s',a'] \Big)$}
			\If {$Q[s,a] = \max_{a}Q(s,a)$}
				\State \algparbox{5.5em}{$\rho \gets (1 - \beta) \rho + \beta \Big(R_a(s,s') - \max_{a}Q[s,a] + \max_{a'}Q[s',a'] \Big)$}
			\EndIf
			\State $s \gets s'$
		\EndFor
	\EndFor
	\State $\pi(s) \gets \text{argmax}_{a} Q[s,a]$ $\forall s \in \mathcal{S}$
	\State \Return $\pi$
\end{algorithmic}
\end{small}	
\end{algorithm}

\section{Simulation Results}
\label{sec_sim}
In this section, we evaluate the performance of the RL approaches in comparison to the greedy policy, where the default action is {\small \textsf{accept}}; and {\small \textsf{federate}} is taken only if the consumer domain has not sufficient resources. The metrics are the average profit \eqref{eq_profit}, and the optimality gap $(AP_{DP} - AP_{Alg}) / AP_{DP}$, where $AP_{Alg}$ is the average profit of the given algorithm. 
In the following, the effect of the capacity of the local and the provider domains and also the impact of the federation cost are investigated. In these simulations, $\gamma = 0.99$ in PI, and $n = 200$, $m = 4000$, $\epsilon = 0.9$, and $\alpha = \beta = 0.9$  in the learning algorithms. The \textit{default} configuration  (if not stated otherwise) of the domains and traffic classes are shown in Table \ref{table_setting}. The following results are the average of 10 experiments.

\begin{table} 
\begin{center}
\centering
\small\addtolength{\tabcolsep}{0pt}
\caption{The Default Simulation Settings}
\label{table_setting}
\scalebox{0.8}{%
\begin{tabular}{|c|c|c|c|c|c|c|c|c|c|c|c|}
\hline
$LC$ & $PC$ & $\lambda_{1}$ & $\mu_{1}$ & $w_{1}$ & $r_{1}$ & $\phi_{1}$ & $\lambda_{2}$ & $\mu_{2}$ & $w_{2}$ & $r_{2}$ & $\phi_{2}$\\
\hline
\hline
30 & 20 & 10 & 4 & 2 & 100 & 30 & 5 & 0.5 & 4 & 20 & 5 \\
\hline 
\end{tabular}	
}		
\end{center}
\end{table}

\subsection{The Effect of the Number of Episodes}
\begin{figure}
\begin{subfigure}[h]{0.49\linewidth}
\includegraphics[width=\linewidth]{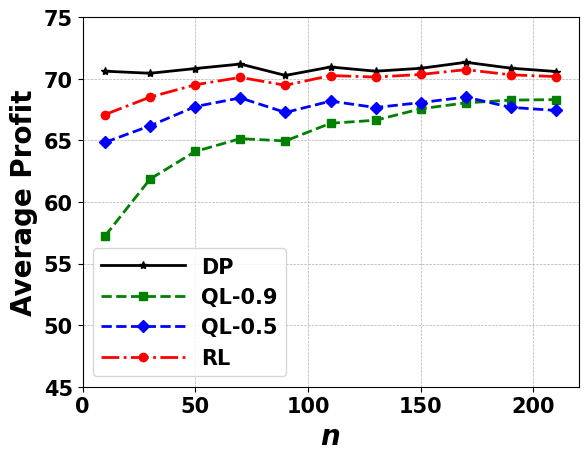}
\caption{Average Profit}
\label{fig_episode_profit}
\end{subfigure}
\hfill
\begin{subfigure}[h]{0.49\linewidth}
\includegraphics[width=\linewidth]{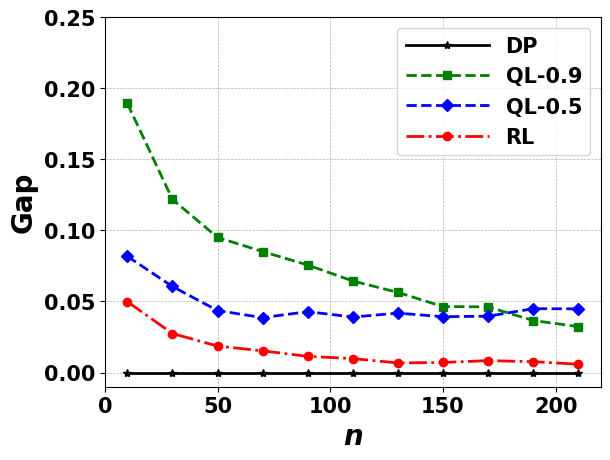}
\caption{Optimality Gap}
\label{fig_episode_gap}
\end{subfigure}
\caption{The effect of the number of the episodes}
\end{figure}

The Q-Learning and R-Learning algorithms learn the policy over time via interaction with the environment. The number of the episodes determines the length of the learning period. The performance of the algorithms with respect to the number of the episodes $n$ is shown in Figures \ref{fig_episode_profit} and \ref{fig_episode_gap}.

The results show that after a number of episodes, both Q-Learning and R-Learning converge. The RL algorithm outperforms QL not only in terms of optimality gap, which is less than 2\%, but also in terms of the number of episodes needed to converge. These results also show the dependency of the performance of Q-Learning on the discount factor $\gamma$. In this setting, while at the beginning, QL with $\gamma = 0.5$ performs better than $\gamma = 0.9$, finally QL-0.9 converges to a better result. 

\subsection{The Effect of the Consumer Domain Capacity}
\begin{figure}
\begin{subfigure}[h]{0.48\linewidth}
\includegraphics[width=\linewidth]{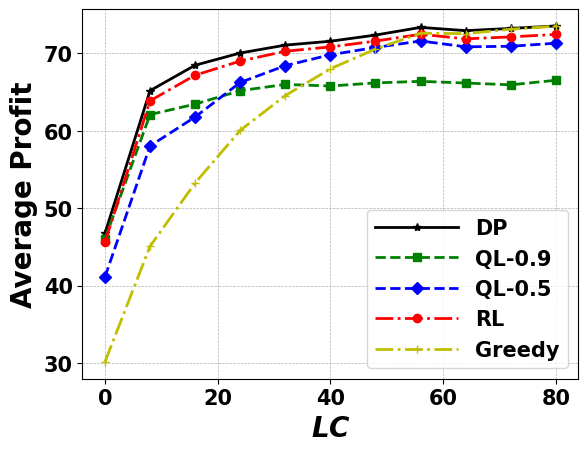}
\caption{Average Profit}
\label{fig_lc_profit}
\end{subfigure}
\hfill
\begin{subfigure}[h]{0.48\linewidth}
\includegraphics[width=\linewidth]{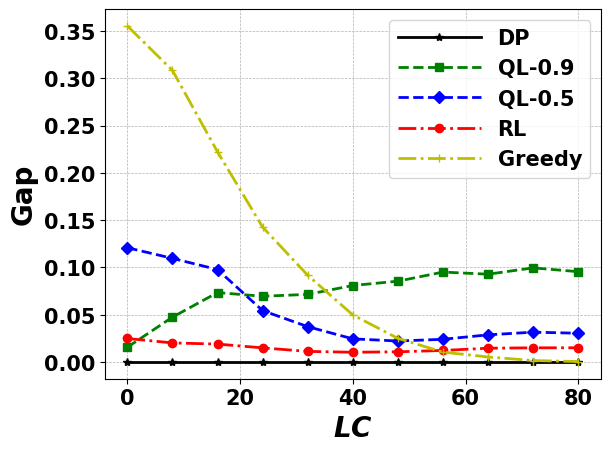}
\caption{Optimality Gap}
\label{fig_lc_gap}
\end{subfigure}
\caption{The effect of the capacity of the consumer domain}
\end{figure}

Efficient management of the local resources of the consumer domain by the AC greatly influences the achievable profit. Figures \ref{fig_lc_profit} and \ref{fig_lc_gap} show the performance of the different policies with respect to the local domain capacity $LC$. {Our simulations show the performances of the algorithms with respect to $PC$ is similar (omitted due to space limitations).} 

The performance of the R-Learning algorithm is excellent independent of $LC$. When the local domain has not sufficient resource to accept all the offered load, i.e., $LC < \sum (w_{i}\lambda_i/\mu_i)$, the intelligent decisions by the RL algorithms greatly improve the performance in comparison to the greedy policy. In the case of very large capacities, all demands can be accepted in the consumer domain, so the greedy algorithm is also the optimal policy. Again, QL performance is affected by $\gamma$ and a single $\gamma$ is not the best choice for all cases. In cases of small capacities, large $\gamma$ is better, but in the large capacities, it should be small.

\subsection{{The Effect of the Offered Load}}
\begin{figure}
\begin{subfigure}[h]{0.48\linewidth}
\includegraphics[width=\linewidth]{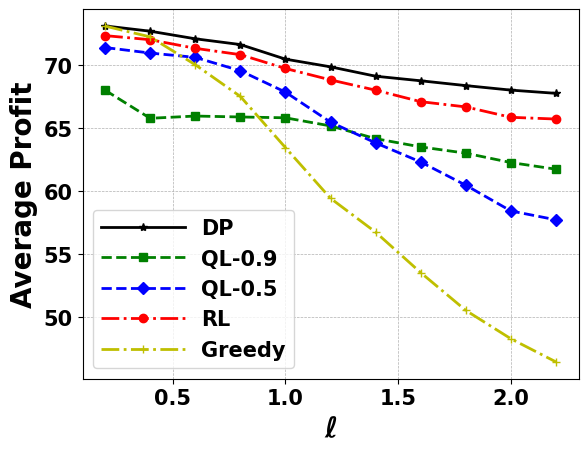}
\caption{Average Profit}
\label{fig_ell_profit}
\end{subfigure}
\hfill
\begin{subfigure}[h]{0.48\linewidth}
\includegraphics[width=\linewidth]{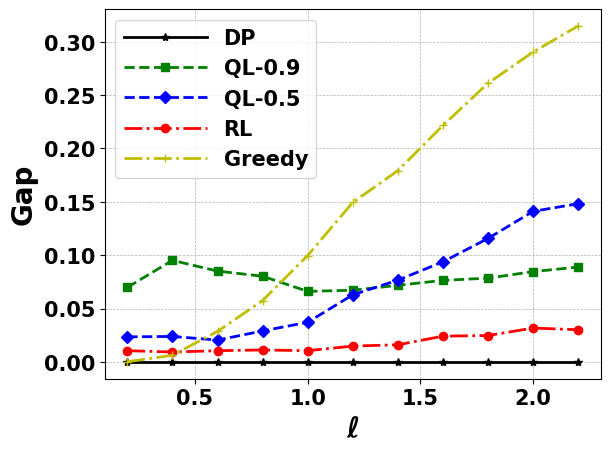}
\caption{Optimality Gap}
\label{fig_ell_gap}
\end{subfigure}
\caption{The effect of the offered load}
\end{figure}

{
The AC, as the resource manager, should efficiently handle the offered load.
In this section, by scaling the arrival rates of demands as $\ell \lambda_{i}$, the performance of the algorithms is investigated. The results are shown in  Figures \ref{fig_ell_profit} and \ref{fig_ell_gap}.

In these  results, it is seen that in the case of very small $\ell$, where all demands can 
be accepted in the consumer domain
as $LC \gg \ell \sum (w_{i}\lambda_i/\mu_i)$,  the greedy policy performs well. However, by increasing the offered load, some demands should be sent to the provider domain or rejected. In these cases, the learning algorithms outperform the greedy policy by making smart decisions. Similar to the other results, the RL algorithm has a near-optimal performance. These results show that the appropriate value of the discount factor $\gamma$ in Q-Learning also depends on the offered load. }

\subsection{The Effect of the Federation Cost}
\begin{figure}
\begin{subfigure}[h]{0.48\linewidth}
\includegraphics[width=\linewidth]{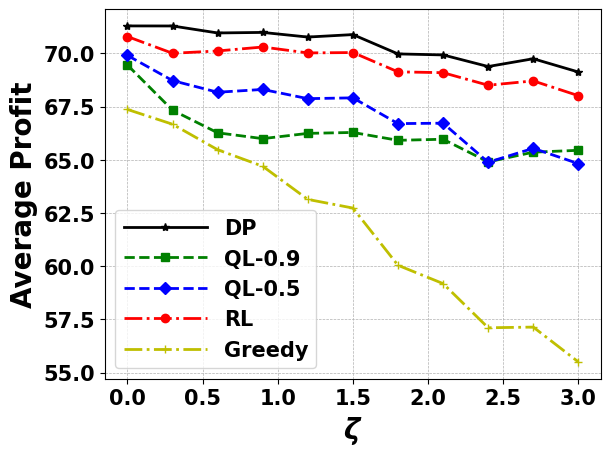}
\caption{Average Profit}
\label{fig_zeta_profit}
\end{subfigure}
\hfill
\begin{subfigure}[h]{0.48\linewidth}
\includegraphics[width=\linewidth]{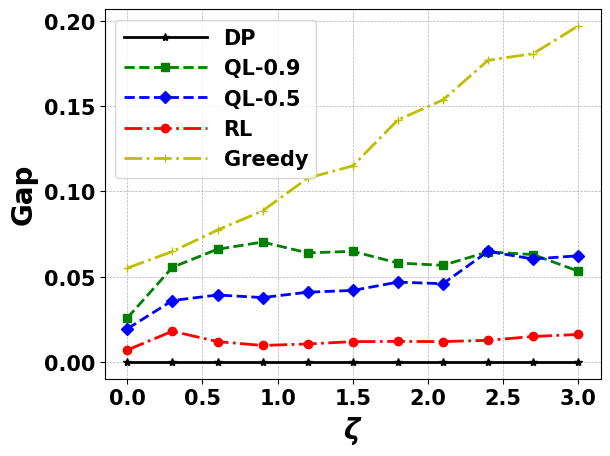}
\caption{Optimality Gap}
\label{fig_zeta_gap}
\end{subfigure}
\caption{The effect of the federation cost}
\end{figure}

As mentioned, for each service type $i$ that is agreed in the federation contract, the provider domain charges the consumer domain an amount $\phi_{i}$ per demand that is deployed in the provider domain. The AC should take this cost into account. For example, if the federation cost is very high, the optimal decision would be to reject the demand. In this section, the admission control policies are evaluated with respect to the cost that is scaled $\zeta \phi_{i}$. Figures \ref{fig_zeta_profit} and \ref{fig_zeta_gap} show the
average profit and the gap of the policies with respect to $\zeta$.  

These results show that by increasing the federation cost, as expected, the profits of all policies decrease. The optimality gap of R-Learning is independent of the scale $\zeta$ that implies it considers the federation cost properly in the admission control process. The Q-Learning algorithm attempts to maintain the gap; however, it does not perform as good as R-Learning. The optimality gap of the greedy policy increases by enlarging $\zeta$ since the non-optimal decisions by the greedy policy incur more cost in the case of larger $\zeta$.



\section{Conclusion and Future Work}
\label{sec_summary}
In this paper, we investigated the ACSF problem, where
the admission controller determines the domain to deploy the service or rejects it in order to maximize the profit considering the federation cost. The optimal policy under the assumption of knowing the arrival and departure rates of the demands was obtained by solving the MDP model of the problem through the policy iteration algorithm. As practical solutions, we applied the Q-Learning and R-Learning algorithms to the problem where the former maximizes the discounted rewards while the latter attempts to maximize the average reward. The extensive simulations show the excellent performance of the R-learning independent of system configuration.

For future work, the next step would be 
taking into account the capacity of the intra-domain links
in multiple provider domains context. These extensions will cause the exponential growth of the state space that needs to be handled by Deep RL solutions.

\section*{Acknowledgment}
We thank Andres Garcia-Saavedra for reviewing and providing feedback on the draft version of the manuscript.

\bibliographystyle{ieeetr}
\bibliography{refs}

\end{document}